# Cryogenic Behavior of High-Permittivity Gate Dielectrics: The Impact of the Atomic Layer Deposition Temperature and the Lithographic Patterning Method

Alessandro Paghi[1*], Sebastiano Battisti[1], Simone Tortorella[1,2], Giorgio De Simoni[1], and Francesco Giazotto[1*]

[1]Istituto Nanoscienze-CNR and Scuola Normale Superiore, Piazza San Silvestro 12, 56127 Pisa, Italy.

[2]Dipartimento di Ingegneria Civile e Industriale, Università di Pisa, Largo Lucio Lazzarino, 56122 Pisa, Italy

[*]Corresponding authors: alessandro.paghi@nano.cnr.it, francesco.giazotto@sns.it

**Abstract**

Dielectrics featuring a high relative permittivity, i.e., high-*k* dielectrics, have become the standard insulators in gate architectures, enhancing the electrical performance of both room temperature and cryogenic electronics. This study delves into the cryogenic (3 K) performance of high-*k* dielectrics commonly used as gate insulators. We fabricated $Al_2O_3$ and $HfO_2$ layers via Atomic Layer Deposition (ALD) and we extrapolated relative permittivity (*k*) and dielectric strength ($E_{BD}$) from AC (100 Hz to 100 kHz) and DC measurements on metal-insulator-metal capacitors. Our findings reveal a strong dependence of $HfO_2$ cryogenic performance on the ALD growth temperature, while the latter shows a negligible impact on $Al_2O_3$. We estimated a ~9 % and ~14 % reduction of the relative permittivity of $HfO_2$ and $Al_2O_3$, respectively, from 300 K to 3 K. Additionally, we designed and fabricated $Al_2O_3/HfO_2$ bilayers and we checked their properties at cryogenic temperatures. The study also investigates the impact of the patterning method, namely, UV or electron-beam lithography (acceleration voltage of 10, 20, or 30 kV), on the high-*k* dielectric properties.

**Introduction**

Cryogenic electronics, which is electronics operating at low temperatures, promises to enhance the performance of electronic systems, from a few transistors to very-large-scale integrated (VLSI) circuits [1][2][3][4][5]. Alongside device miniaturization, lowering the operation temperature promises an advancement in performance thanks to the improvement in materials-related properties [2]. Increasing the carrier mobility and lowering the resistance, namely down to 0 Ω in the case of superconductors [6], allows to reach higher operational speeds decreasing the temperature [7]. At the same time, higher conductivities, together with the reduction of current leakages, (sub)thresholds, and supply voltages, allow to break down power requirements [8][9].

In the same manner of room temperature electronics, Metal Oxide Semiconductor Field Effect Transistors (MOS-FETs) working at cryogenic temperatures, i.e., cryo-CMOS, are considered the building block for most of the low-temperature electrical applications [10][11][12]. Specifically, due to the previously mentioned achievements, electrical performance of MOSFETs tend to improve by lowering the device temperature, reaching higher gain and speed, lower noise level, and improved reliability [13]. Similar considerations could be done if superconductive source and drain leads are included in the FET, leading to a superconducting Josephson Field Effect Transistor (JoFET) [14][15][16][17]. The latter device is an essential cryogenic building block for quantum fundamental research activities [18][19][20] and technological applications [21][22][23][24]. An example of this is the gatemon qubit [25][26][27][28][29]. The gatemon is derived from the transmon qubit, a resonant superconducting circuit that relies on the nonlinear inductance of Josephson Junctions [30][31][32]. The gatemon allows an easy tuning of the qubit transition frequency by using the gate voltage to change the inductance of a JoFET [25][26][27][28][29]. In contrast to the flux-controlled transmon, the voltage-tunable gatemon shows promise in simplifying the scaling-down of multi-qubit circuits and provide new methods to control multi-qubit architectures.

Starting from the 45-nm technological node, dielectrics featuring a high relative permittivity, i.e., high-$k$ dielectrics, are the standard insulators embedded in gate architectures to decrease the gate voltage operation range of a FET [33][34][35]. Among these, commonly used high-$k$ dielectrics are, but not limited to, Aluminum Oxide ($Al_2O_3$, $k$~9 [36][37][38][39]), Hafnium Dioxide ($HfO_2$, $k$~25 [36][37][38][39]), Zirconium Oxide ($ZrO_2$, $k$~25 [36][37][38][40]), Lanthanum Oxide ($La_2O_3$, $k$~30 [36][37][38][41]), and Titanium Dioxide ($TiO_2$, $k$~80 [36][37][38][42]). Consequently, it is essential to characterize the cryogenic properties of high-*k* dielectrics to improve classic and superconducting FET design at cryogenic temperatures [43].

In this study, we assessed the cryogenic (3 K) performance of high-*k* dielectrics commonly used as gate insulators. Specifically, we examined relative permittivity (*k*) and dielectric strength ($E_{BD}$) of $Al_2O_3$ and $HfO_2$ layers fabricated using Atomic Layer Deposition (ALD). We determined the dielectric properties through AC (100 Hz to 100 kHz) and DC measurements on metal-insulator-metal capacitors. Our results show a ~9% reduction in the relative permittivity of $HfO_2$ and a ~14% reduction in $Al_2O_3$, as the temperature decreases from 300 K to 3 K. We observed that the cryogenic performance of $HfO_2$ is greatly influenced by the ALD growth temperature, while the latter shows a negligible impact on $Al_2O_3$. We also fabricated $Al_2O_3$/$HfO_2$ bilayers, and we analyzed their properties at cryogenic temperatures. Furthermore, we investigated the impact of different patterning methods, such as UV or electron-beam lithography, on the properties of high-*k* dielectrics.

**Experimental Section**

*Fabrication of MIM Capacitors*

Metal-insulator-metal (MIM) capacitors were fabricated on $SiO_2$(300nm-thick)/Si square samples. "T-shape" bottom electrodes were obtained via a Ti/Au(10/40 nm) bilayer lift-off. Then, single layers of $HfO_2$ or $Al_2O_3$, and a bilayer of $Al_2O_3$/ $HfO_2$ were deposited via Atomic Layer Deposition (ALD) at an optimized temperature of 130 °C or 300 °C. For $HfO_2$, Tetrakis(ethylmethylamino)hafnium (TEMAH) and $H_2O$ were used as oxide precursors, while Trimethylaluminum (TMAl) and $H_2O$ were involved in the $Al_2O_3$ growth. In both cases, Ar was used as carrier gas. For both the insulators, the deposition process follows 4 steps: (i) main precursor dose, (ii) main precursor purge, (iii) $H_2O$ dose, and (iv) $H_2O$ purge. More details about the ALD steps are provided in the Supporting Information. Single layers were deposited through 250 ALD cycles of $HfO_2$ or $Al_2O_3$; on the other hand, the bilayers were deposited through 125 ALD cycles of $Al_2O_3$ followed by 125 ALD cycles of $HfO_2$ grown at the same temperature.
After the insulator deposition, aligned "T-shape" top electrodes were obtained via a Ti/Au(10/55 nm) bilayer lift-off. To check the impact of the patterning method on the electrical performance of the deposited insulators, the top electrode patterning method was chosen between UV lithography and electron beam lithography (EBL). The EBL was performed at an acceleration voltage of 10 kV, 20 kV, or 30 kV, with gun apertures of 7.5 μm or 120 μm.

*Electrical Characterization of MIM Capacitors*

AC electrical characterization of MIM capacitors was performed by measuring the capacitor impedance (magnitude and phase) in the frequency range from 100 Hz to 100 kHz upon application of a sinusoidal voltage signal with 250 mV$_{rms}$ amplitude and 0 V bias. DC electrical characterization of MIM capacitors was performed by recording the current-voltage curve upon application of a step-graded voltage (absolute voltage step = 100 mV, step delay = 1 s) and measuring the flowing current.

An extended version of the experimental section containing information about "Materials and chemicals", "Fabrication of MIM capacitors", "Electrical and morphological characterization of MIM capacitors", "Best-fitting of experimental data", and "Statistical Analysis" is provided in the Supporting Information.

**Results and Discussion**

Figure 1a shows the 3D model of a T-shaped MIM capacitor used to evaluate the electrical performance of high-k dielectrics.

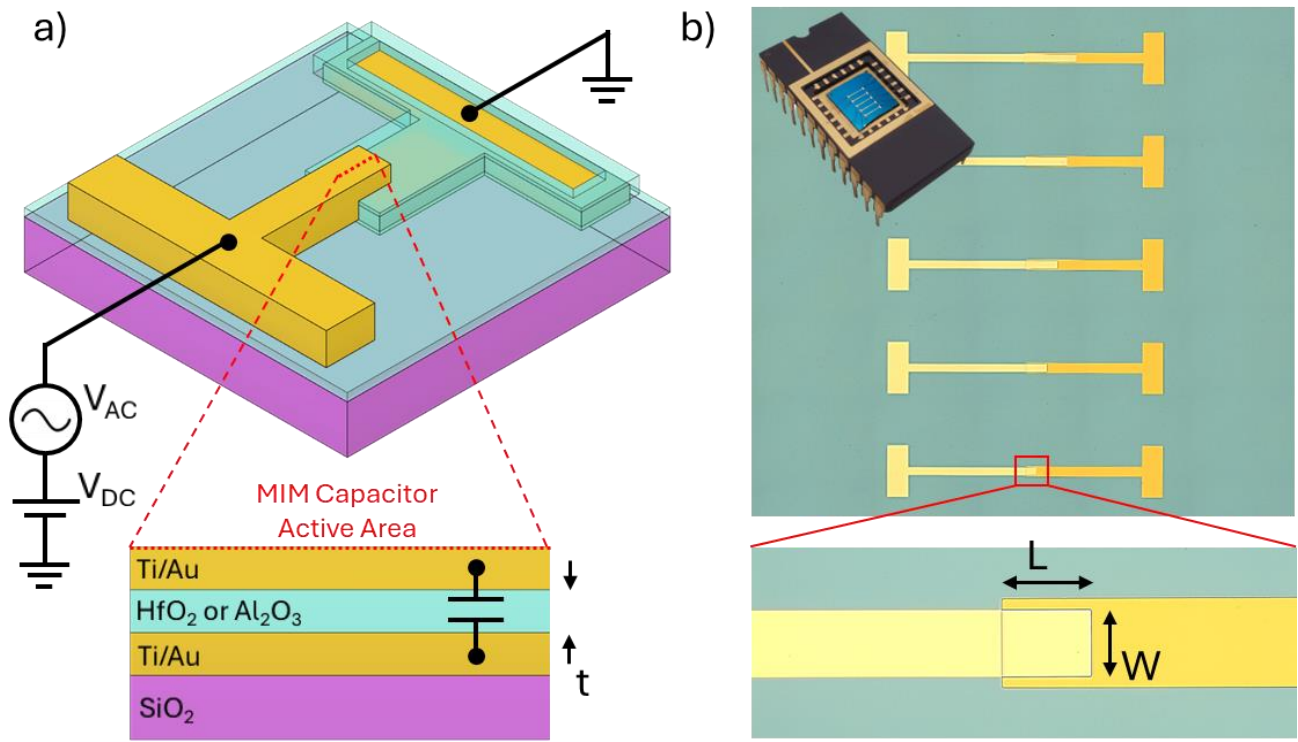

**Figure 1: Metal-Insulator-Metal capacitors: concept and morphologic characterization.** a) 3D sketch of a MIM capacitor used to measure electrical performance of high-$k$ dielectrics. The inset shows the in-section MIM capacitor active area where the high-$k$ dielectric acts as the primary insulator. b) Optical microscopy image of a chip featuring MIM capacitors with different areas and HfO$_2$ as an insulator; the scalebar is 1 mm. The inset shows the MIM capacitor active area; the scalebar is 100 μm.

Briefly, a Ti/Au (10/40 nm thick) "T shape" bottom metal electrode was deposited via thermal evaporation on a SiO$_2$/Si substrate. Then, a high-*k* dielectric layer, which is HfO$_2$, Al$_2$O$_3$, or a Al$_2$O$_3$/HfO$_2$ bilayer, was conformably deposited via Atomic Layer Deposition (ALD). We investigated two ALD deposition temperatures, namely 130 °C and 300 °C. Single layers were deposited through 250 ALD cycles, while the Al$_2$O$_3$/HfO$_2$ bilayer via 125 cycles of Al$_2$O$_3$ followed by 125 cycles of HfO$_2$. The resulting insulating thickness is ~ 31 nm and ~ 28 nm (growth rate of 1.25 Å/cycle and 1.10 Å /cycle) for insulators grown at 130 °C and 300 °C, respectively. Eventually, an aligned Ti/Au (10/55 nm thick) "T shape" top metal electrode was deposited via thermal evaporation to overlap part of the bottom electrode. The overlapping area (A=W×L), which was set to 7500 μm$^2$, 15000 μm$^2$, 22500 μm$^2$, 30000 μm$^2$, or 37500 μm$^2$, is the MIM capacitor active area where the high-*k* dielectric acts as the main insulator. Figure 1b shows an optical microscopy image of a 7×7 mm$^2$ die embedding MIM capacitors. As observed from the photograph, the die contains 5 capacitors where the overlapping area increases from bottom to top.

MIM capacitors featuring HfO$_2$ or Al$_2$O$_3$ as gate insulators were measured in a $^4$He closed-cycle cryostat with a low-frequency measurement setup; the electrical characterization was performed at 300 K and 3K. We measured the capacitor impedance, namely, magnitude (|Z|) and phase (∠Z), over the frequency range from 100 Hz to 100 kHz to assess the performance of MIM capacitors [39][44]. Figures 2a,d show impedance magnitude and phase over frequency, i.e., the Bode diagrams, of MIM capacitors featuring HfO$_2$ (Figure 2a) or Al$_2$O$_3$ (Figure 2d) as insulators, measured at 3 K.

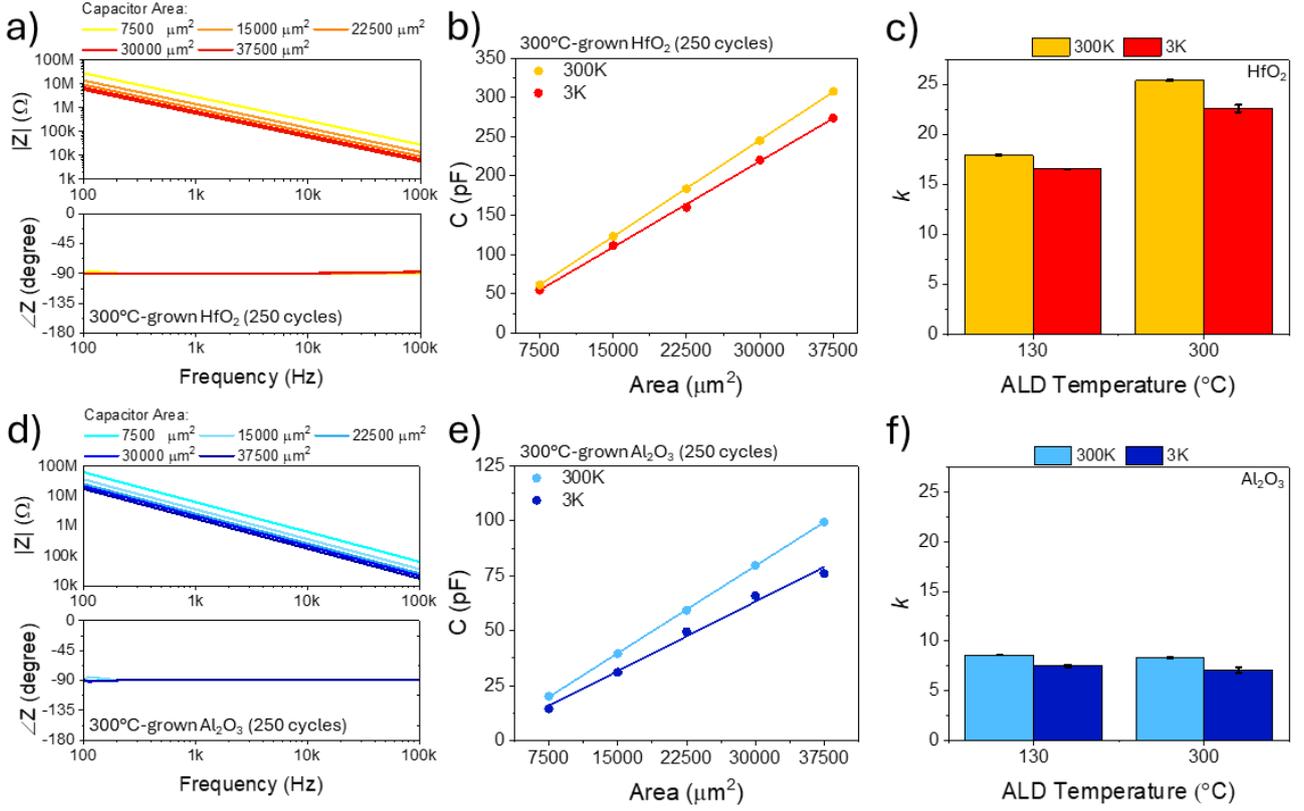

**Figure 2: HfO₂ and Al₂O₃ cryogenic relative permittivity behavior.** a,d) Bode diagrams of MIM capacitors featuring $HfO_2$ (a) or $Al_2O_3$ (d) fabricated at an ALD temperature of 300 °C, measured at 3 K. b,e) Capacitance vs. capacitor area of MIM capacitors featuring $HfO_2$ (b) or $Al_2O_3$ (e) fabricated at an ALD temperature of 300 °C, measured at 300 K and 3 K. c,f) Dielectric permittivity of high-$k$ insulators grown via ALD, measured at 300 K and 3 K. Data in (c,f) are reported as the average value measured over 5 MIM capacitors, with error bars representing the root mean square deviation.

The impedance behavior follows what is expected for an ideal capacitor: the magnitude of the impedance depends on frequency (f) and capacitance (C) as $1/2\pi fC$, while the phase exhibits a constant value of ~-90°. Through best-fitting ($R^2 = 0.9999$) of experimental data on magnitude and phase vs. frequency with an electrical lumped model consisting of a capacitor C, we retrieved the capacitance of MIM capacitors:

$$(1) \; C = \varepsilon_0 k \frac{A}{t},$$

where $\varepsilon_0$ is the vacuum permittivity, $k$ is the relative permittivity, A is the capacitor area, and $t$ is the dielectric thickness. Figure 2b,e show capacitance values of MIM capacitors featuring $HfO_2$ (Figure 2b) or $Al_2O_3$ (Figure 2e) grown via ALD at 300 °C, measured both at 300 K and 3 K. The capacitance linearly increases with the capacitor area, in agreement with (1). For each area, the capacitance value decreases by reducing the temperature from 300 K to 3 K. We then extrapolated the relative permittivity by inverting (1), under the approximation of negligible changes of in-plane and out-of-plane MIM capacitor geometric features with the temperature [45]. Figure 2c,f show $k$ values of HfO$_2$ (Figure 2c) and Al$_2$O$_3$ (Figure 2f) insulators grown via ALD at 130 °C and 300 °C, measured both at 300 K and 3 K. For all the insulators, the relative permittivity decreases by reducing the device temperature. An average decrement of ~9 % and ~14 % was evaluated for HfO$_2$ and Al$_2$O$_3$, respectively, which agrees with our recent findings with other popular conventional insulators [46]. Interestingly, Al$_2$O$_3$ shows a relative permittivity of ~8.5 and ~7.3 at 300 K and 3 K, respectively, regardless of the ALD growth temperature. On the other hand, a strong impact of the ALD deposition parameters is observed on the HfO$_2$ properties. Specifically, in the case of the 130°C-grown HfO$_2$, the layer features a $k$ of 17.9 and 16.5 at 300 K and 3 K, while in the case of the 300°C-grown HfO$_2$, the insulator shows a $k$ of 25.4 and 22.6 at 300 K and 3 K, respectively. We associated the lower permittivity of the 130°C-grown HfO$_2$ to a reduced film density induced by the lower ALD growth temperature [47].

Next, we measured the MIM capacitor current vs. voltage curves, i.e., I-V curves, from which the avalanche breakdown behavior of the fabricated insulators is observed. Figure 3a,c show I-V curves of MIM capacitors featuring HfO$_2$ (Figure 3a) or Al$_2$O$_3$ (Figure 3c) as insulators, measured at 3 K.

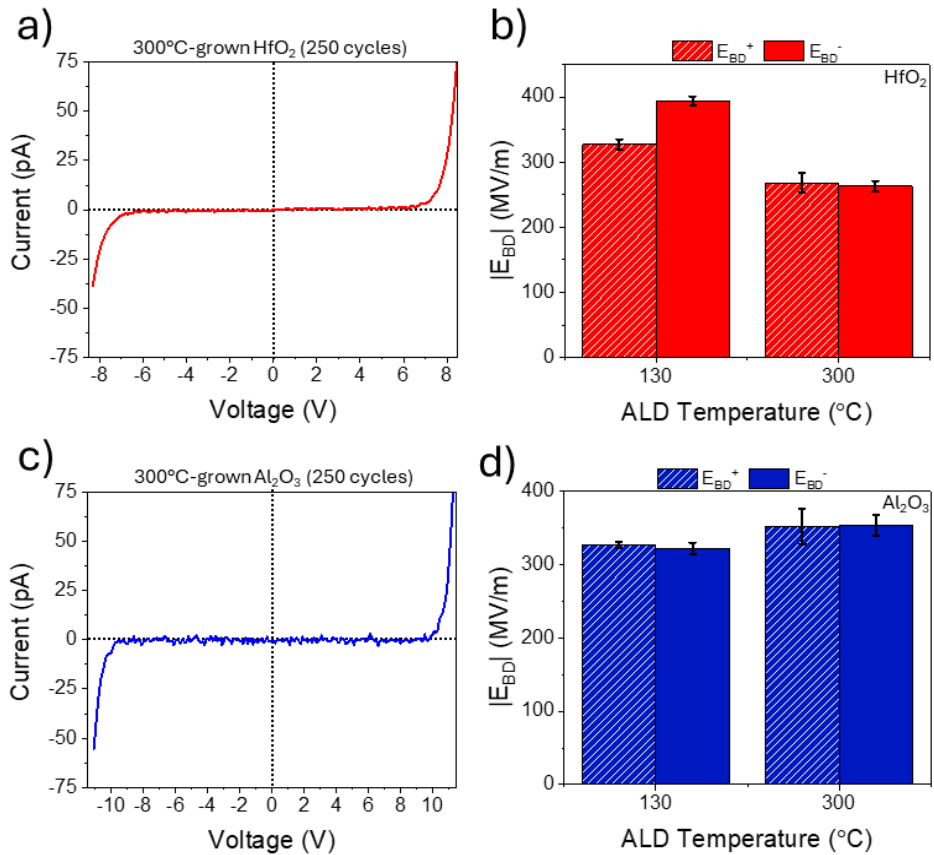

**Figure 3: HfO$_2$ and Al$_2$O$_3$ cryogenic dielectric strength behavior.** a,c) Current vs. voltage curve of MIM capacitors featuring HfO$_2$ (a) or Al$_2$O$_3$ (c) fabricated at an ALD temperature of 300 °C,

measured at 3K. b,d) Absolute value of positive and negative dielectric strengths of high-*k* insulators grown via ALD, measured at 3 K. Data in (b,d) are reported as the average value measured over 5 MIM capacitors, with error bars representing the root mean square deviation.

The current slowly linearly increases with the voltage applied until it reaches the breakdown voltage ($V_{BD}$), where the relationship between current and voltage becomes exponential. We best-fitted ($R^2$ = 0.97) experimental data with a two-regions piecewise linear model to retrieve positive ($V_{BD}^+$) and negative ($V_{BD}^-$) breakdown voltages (Figure S1). Then, the dielectric strength ($E_{BD}$) was extrapolated as:

$$(2)\ E_{BD} = -\frac{V_{BD}}{t}.$$

Figure 3b,d show $E_{BD}$ values of $HfO_2$ (Figure 3b) and $Al_2O_3$ (Figure 3d) grown via ALD at 130 °C and 300 °C, measured at 3 K. Apart for the $HfO_2$ layer grown at 130 °C, all the dielectrics exhibit similar positive and negative dielectric strengths. As reported for *k*, $Al_2O_3$ shows a dielectric strength of ~330 MV/m, regardless of the ALD growth temperature. On the other hand, while the 300°C-grown $HfO_2$ features a symmetric $|E_{BD}|$ of ~270 MV/m, the 130°C-grown $HfO_2$ shows an asymmetric dielectric strength where $E_{BD}^+$ = -330 MV/m and $E_{BD}^-$ = 400 MV/m. The asymmetric behavior observed for the $HfO_2$ layer grown at 130 °C could be related to the presence of polarized chemical and electrical impurities incorporated during the low temperature ALD process [47][48][49].

We then manufactured MIM capacitors featuring a $Al_2O_3/HfO_2$ bilayer, which is fabricated with 125 ALD cycles of $Al_2O_3$ followed by 125 ALD cycles of $HfO_2$. Bilayers of high-*k* dielectrics are helpful in all those situations where low chemical compatibility between metallic or semiconductive electrodes and insulators is present. An example of this is the $InAs/HfO_2$ case, where the issue can be circumvented by interposing a thin layer of $Al_2O_3$ between them [50]. Figure 4a,b show capacitance values of MIM capacitors featuring a bilayer of $Al_2O_3/HfO_2$ grown via ALD at 130 °C (Figure 4a) or 300 °C (Figure 4b), measured at 3 K.

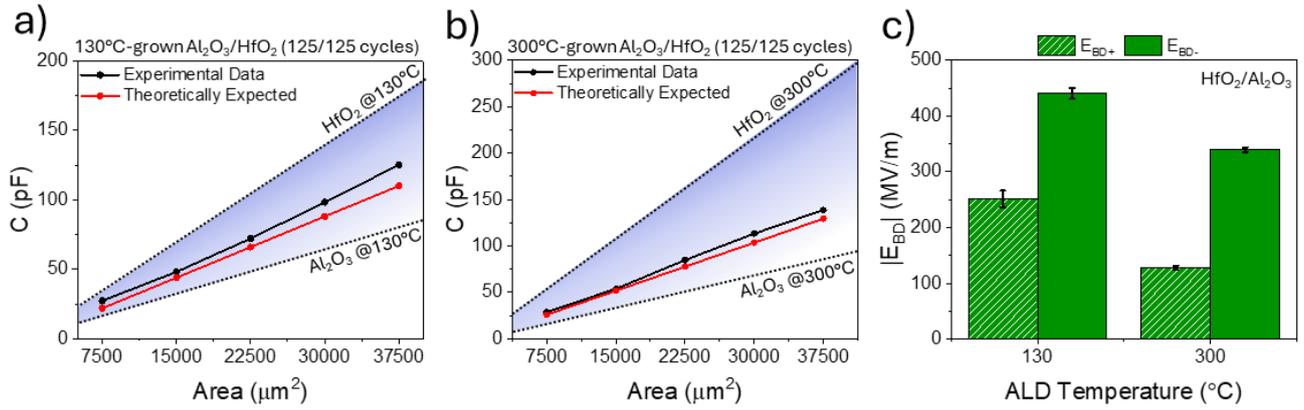

**Figure 4: Cryogenic behavior of Al$_2$O$_3$/HfO$_2$ bilayers.** a,b) Capacitance vs. capacitor area of MIM capacitors featuring a bilayer of Al$_2$O$_3$/HfO$_2$ grown at 130 °C (a) or 300 °C (b), measured at 3K. c) Absolute value of positive and negative dielectric strengths of Al$_2$O$_3$/HfO$_2$ bilayers grown via ALD, measured at 3 K. Data in (c) are reported as the average value measured over 5 MIM capacitors, with error bars representing the root mean square deviation.

Capacitance experimental data (black dots) are in good agreement with what is expected from the theoretical model (red dots):

$$(3) \quad \frac{1}{C_{HfO_2/Al_2O_3}} = \frac{1}{C_{HfO_2}} + \frac{1}{C_{Al_2O_3}},$$

where C are calculated from (1). Values of relative permittivities involved in calculations were retrieved from Figure 2c,f, while thicknesses are obtained dividing by 2 the measured thickness of HfO$_2$ and Al$_2$O$_3$ layers. Figure 4c shows $E_{BD}$ values of HfO$_2$/Al$_2$O$_3$ bilayers, measured at 3 K. For both the ALD growth temperatures, the dielectrics exhibit an asymmetric dielectric strength, also if the starting layers exhibit symmetrical values. This could be related to the MIM capacitor asymmetry in the out of plane direction with respect to the applied voltage drop: both band gaps of HfO$_2$ and Al$_2$O$_3$ and work functions of Ti (in contact with HfO$_2$) and Au (in contact with Al$_2$O$_3$) are different. This kind of structure produces an asymmetrical behavior in the transmission of electrons through the bilayer, which could result in an asymmetrical dielectric strength [51][52].

Eventually, we evaluated the impact of the patterning method used to define the MIM capacitor top electrode geometry on the cryogenic properties of high-*k* dielectrics. As a case of study, we focused on HfO$_2$ grown via ALD at 130 °C and we explored different patterning techniques, namely, UV lithography (UVL) and electron-beam lithography (EBL) (Figure 5a). EBL was performed changing the acceleration voltage (V) from 10 kV to 30 kV and the gun aperture (ϕ) from 7.5 μm to 120 μm. The investigation was triggered since irradiation of semiconductor or insulators with high energy electrons is able to damage the material, drastically changing its transport properties [53][54]. Figure S2 shows simulated electrons trajectories and deposited energy for irradiating electrons with V of 10, 20, 30 kV. Regardless of the accelerating voltage, in all cases irradiating electrons reach the HfO$_2$ layer with an energy density comparable with that released in the EBL resist.

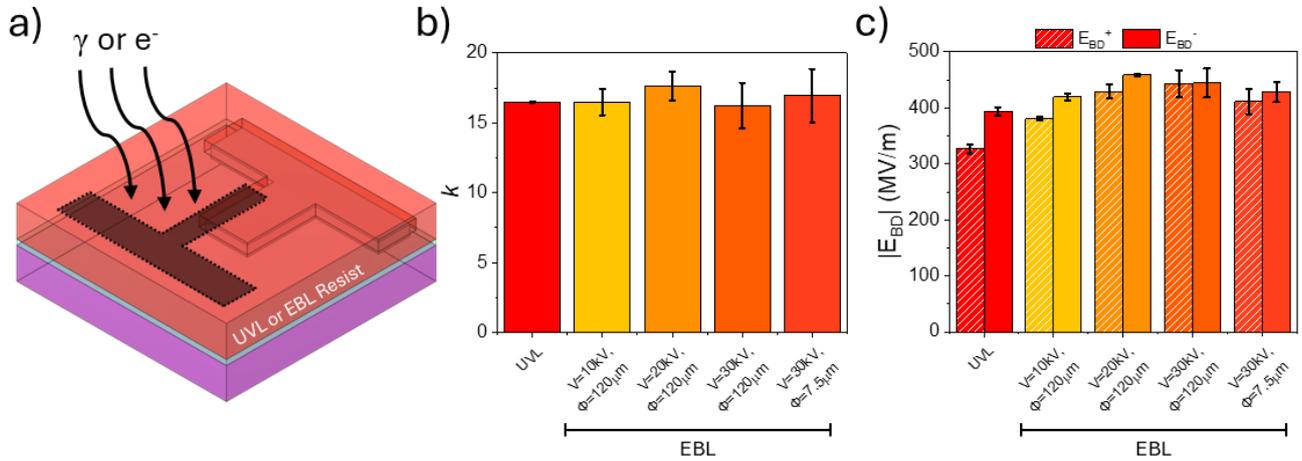

**Figure 5: Impact of the lithography patterning method on the HfO$_2$ cryogenic dielectric behavior.** a) 3D sketch of the MIM capacitor top electrode lithography patterning method. UV lithography or electron-beam lithography are involved as patterning techniques. b) Dielectric permittivity of HfO$_2$ grown via ALD at 130 °C changing the patterning technique, measured at 3 K. b) Absolute value of positive and negative dielectric strengths of HfO$_2$ grown via ALD at 130 °C changing the patterning technique, measured at 3 K. Data in (b,c) are reported as the average value measured over 5 MIM capacitors, with error bars representing the root mean square deviation.

Figure 5b shows relative permittivity values of the 130°C-grown HfO$_2$ layer measured at 3K, from which no significant variations are observed changing the lithographic method and settings. The same is true for the dielectric strength (Figure 5c), where the asymmetric behavior of the 130°C-grown HfO$_2$ layer was substantially maintained for all the patterning processes developed. The results support the statement that irradiation of high-*k* dielectrics during EBL with electrons featuring typical EBL doses and energies ranging from 10 keV to 30 keV does not alter their cryogenic dielectric behavior.

**Conclusions**

In this work, we studied the behavior of high-*k* dielectrics at cryogenic temperatures (3 K). We fabricated $Al_2O_3$ and $HfO_2$ layers via Atomic Layer Deposition (ALD) and we extrapolated relative permittivity (*k*) and dielectric strength ($E_{BD}$) from AC (100 Hz to 100 kHz) and DC measurements on metal-insulator-metal capacitors. $Al_2O_3$ showed consistent cryogenic performance regardless of the ALD growth temperature, while the properties of $HfO_2$ strongly depended on the ALD parameters. We observed approximately a 9 % and 14 % reduction of the relative permittivity of $HfO_2$ and $Al_2O_3$, respectively, from 300 K to 3 K. Additionally, we fabricated $Al_2O_3$/$HfO_2$ bilayers which can be employed as gate dielectric stacks in specific cryogenic applications. We also investigate the effect of different patterning methods, such us UV or electron-beam lithography (EBL), on the properties of high-*k* dielectrics. The results demonstrated that the irradiation of high-*k* dielectrics during EBL with electrons featuring typical EBL doses and energies ranging from 10 keV to 30 keV does not change their cryogenic dielectric behavior.

**Supporting Information**

The supporting information document contains the supplementary Figures S1 and S2, reporting best-fitting curves of I-V characteristics for $V_{BD}$ estimation and trajectories of irradiating electrons during EBL, respectively. The supporting information also includes additional experimental details, materials, and methods.

**Acknowledgments**

This work was supported in part by EU's Horizon 2020 Research and Innovation Framework Program under Grant 964398 (SUPERGATE), by Grant 101057977 (SPECTRUM), and in part by the Piano Nazionale di Ripresa e Resilienza, Ministero dell'Università e della Ricerca (PNRR MUR) Project under Grant PE0000023-NQSTI.

**Supporting Information**

**Cryogenic Behavior of High-Permittivity Gate Dielectrics: The Impact of the Atomic Layer Deposition Temperature and the Lithographic Patterning Method**

Alessandro Paghi[1*], Sebastiano Battisti[1], Simone Tortorella[1,2], Giorgio De Simoni[1], and Francesco Giazotto[1*]

[1]Istituto Nanoscienze-CNR and Scuola Normale Superiore, Piazza San Silvestro 12, 56127 Pisa, Italy.

[2]Dipartimento di Ingegneria Civile e Industriale, Università di Pisa, Largo Lucio Lazzarino, 56122 Pisa, Italy

[*]Corresponding authors: alessandro.paghi@nano.cnr.it, francesco.giazotto@sns.it

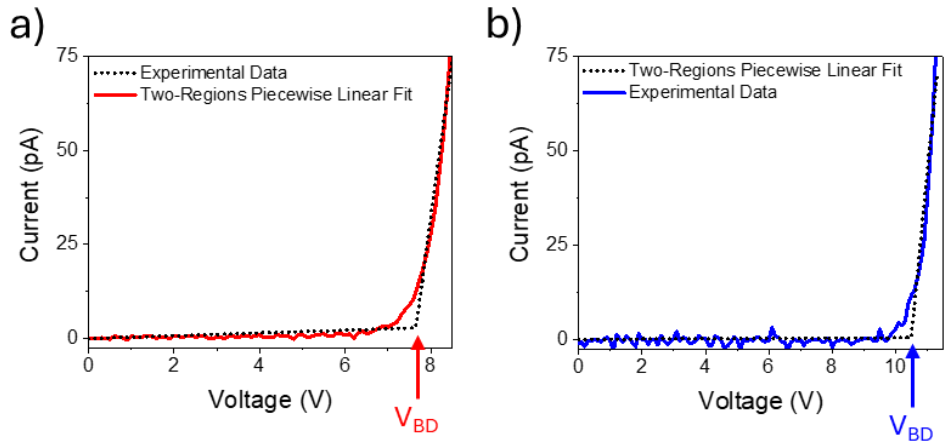

**Figure S1: Two-regions piecewise linear best fitting of current vs. voltage curve of MIM capacitors.** a,b) Experimental and best-fitted current vs. voltage curves of MIM capacitors featuring $HfO_2$ (a) or $Al_2O_3$ (c) fabricated at an ALD temperature of 300 °C, measured at 3K.

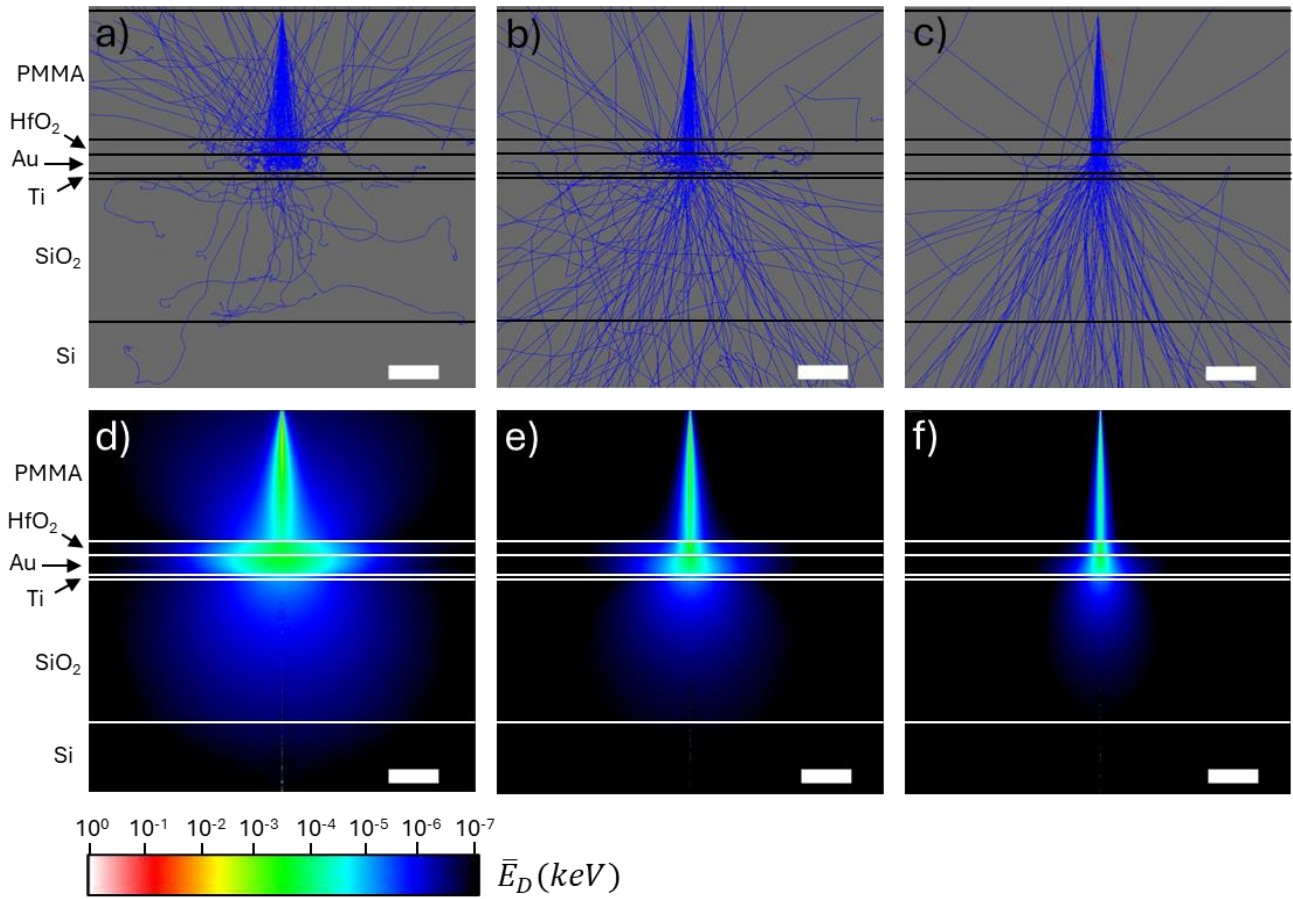

**Figure S2: Electron Beam Lithography simulated electrons trajectories and average deposited energy.** a,b,c) Trajectories of irradiating (blue) and secondary (red) electrons with an accelerating voltage of 10 kV (a), 20 kV (b), and 30 kV (c). d,e,f) Average deposited energy by irradiating

electrons with an accelerating voltage of 10 kV (d), 20 kV (e), and 30 kV (f). The scalebar is 100 nm.

## Materials and Methods

### Materials and Chemicals

SiO$_2$(300-nm-thick)/Si wafers (4'' diameter, (100) orientation, thickness 400 μm, ρ > 10000 Ω×m) were purchased from Siltronix. Acetone (ACE, ULSI electric grade, MicroChemicals), 2-propanol (IPA, ULSI electric grade, MicroChemicals), LOR 3A (Kayaku AM). S1805 G2 Positive Photoresist (S1805, Microposit, positive photoresist), AR-P 679.04 (AllResist, positive e-beam resist), MF319 Developer (MF319, Microposit), AR 600-56 Developer (AR 600-56, AllResist), AR600-71 (AllResist, remover for photo- and e-beam resist), Nitrogen (N$_2$, 5.0, Nippon Gases) were provided by the Clean Room Facility of the National Enterprise for nanoScience and nanotechnology (NEST, Pisa, Italy). Aqueous solutions were prepared using deionized water (DIW, 15.0 MΩ×cm) filtered by Elix® (Merck Millipore) provided by the Clean Room Facility of the NEST.

### Fabrication of Metal-Insulator-Metal (MIM) Capacitors

All wetting steps were performed in cleaned glass Beckers using stainless steel tweezers provided with carbon tips. Teflon-coated tweezers were used for all the steps requiring acid or base solutions. The Julabo TW2 was used to head the solution to a specific temperature. Unless otherwise stated, all the rinsing steps were performed at room temperature (RT, 21 °C).

SiO$_2$(300nm-thick)/Si wafers were cut in square samples (7×7 mm×mm) and sonicated (Transonic, T310/H) in ACE and IPA for 5 min to remove Si dusts and organic contaminants. Samples were then subjected to a O$_2$ plasma cleaning (Electronic Dienier, Femto, residual pressure = 25 mTorr, plasma pressure = 70 mTorr, power = 100 W) for 90 s. After that, a LOR3A/S1805 bilayer was deposited. LOR3A was spin-coated at 4000 RPM for 60 s (spin coating acceleration of 4000 RPM/s) and soft-baked at 170 °C for 180 s, while S1805 positive photoresist was spin-coated at 5000 RPM for 60 s (spin coating acceleration of 5000 RPM/s) and soft-baked at 115 °C for 60 s. The resist was then exposed via direct writing UV lithography (UVL, DMO, ML3 laser writer, λ=385 nm) with a dose of 45 mJcm$^{-2}$, resolution of 0.6 μm, high exposure quality, and laser-assisted real-time focus correction to define the bottom electrode geometry. The UV-exposed samples were developed in MF319 for 120 s with soft agitation to remove exposed S1805 and the underlying layer of LOR3A, then rinsed in DIW for 30 s to stop the development and dried with N$_2$. The photoresist-patterned samples were loaded into a thermal evaporator where a 10/40-nm-thick Ti/Au bilayer was deposited at a rate of 1.5 A/s at a residual chamber pressure of 2E-6 mbar. The deposited film was lifted-off in AR600-71 at 80 °C for 5 min with strong agitation, then rinsed in IPA for 30 s, and dried with $N_2$, to obtain the metallic bottom electrodes.

Samples were loaded into the vacuum chamber of an Atomic Layer Deposition system (ALD, Oxford Instruments, OpAL) where the high-$k$ dielectric was uniformly grown at 130 °C or 300 °C. We deposited single layers of $HfO_2$ and $Al_2O_3$, and a bilayer of $Al_2O_3$/ $HfO_2$. For $HfO_2$, Tetrakis(ethylmethylamino)hafnium (TEMAH) and $H_2O$ were used as oxide precursors, while Trimethylaluminum (TMAl) and $H_2O$ were involved in the $Al_2O_3$ growth. In both the cases, Ar was used as carrier gas. Ar bubbling was also involved to increase the volatility of TEMAH. After reaching a base pressure of ~3-4 mTorr, the chamber pressure was increased to ~360 mTorr injecting Ar. For both the insulators, the deposition process follows 4 steps: (i) main precursor dose, (ii) main precursor purge, (iii) $H_2O$ dose, and (iv) $H_2O$ purge.

For $HfO_2$ grown at an ALD temperature of 130 °C:
    (i)    TEMAH dose: TEMAH valve on; Ar bubbler: 250 sccm; Ar purge: 10 sccm; step duration: 0.9 s.
    (ii)    TEMAH: purge: TEMAH valve off; Ar purge: 250 sccm; step duration: 110 s.
    (iii)    $H_2O$ dose: $H_2O$ valve on; Ar purge: 10 sccm; step duration: 0.03 s.
    (iv)    $H_2O$ purge: $H_2O$ valve off; Ar purge: 250 sccm; step duration: 90 s.

For $HfO_2$ grown at an ALD temperature of 300 °C:
    (i)    TEMAH dose: TEMAH valve on; Ar bubbler: 250 sccm; Ar purge: 10 sccm; step duration: 0.6 s.
    (ii)    TEMAH: purge: TEMAH valve off; Ar purge: 300 sccm; step duration: 12 s.
    (iii)    $H_2O$ dose: $H_2O$ valve on; Ar purge: 250 sccm; step duration: 0.02 s.
    (iv)    $H_2O$ purge: $H_2O$ valve off; Ar purge: 250 sccm; step duration: 20 s.

For $Al_2O_3$ grown at an ALD temperature of 130 °C:
    (i)    TMAl dose: TMAl valve on; Ar purge: 50 sccm; step duration: 0.03 s.
    (ii)    TMAl purge: TMAl valve off; Ar purge: 250 sccm; step duration: 110 s.
    (iii)    $H_2O$ dose: $H_2O$ valve on; Ar purge: 5 sccm; step duration: 0.025 s.
    (iv)    $H_2O$ purge: $H_2O$ valve off; Ar purge: 250 sccm; step duration: 90 s.

For $Al_2O_3$ grown at an ALD temperature of 300 °C:
    (i)    TMAl dose: TMAl valve on; Ar purge: 50 sccm; step duration: 0.02 s.
    (ii)    TMAl purge: TMAl valve off; Ar purge: 100 sccm; step duration: 6 s.

(iii)      $H_2O$ dose: $H_2O$ valve on; Ar purge: 5 sccm; step duration: 0.020 s.

(iv)      $H_2O$ purge: $H_2O$ valve off; Ar purge: 100 sccm; step duration: 12 s.

Single layers were deposited through 250 ALD cycles of $HfO_2$ or $Al_2O_3$; on the other hand, the bilayers were deposited through 125 ALD cycles of $Al_2O_3$ followed by 125 ALD cycles of $HfO_2$ grown at the same temperature. Eventually, the chamber was pumped to reach a base pressure of ~3-4 mTorr and then vented using $N_2$.

After the insulator deposition, samples were removed from the ALD system and heated up to 170 °C for 180 s to desorb adsorbed water molecules from air. After that, for samples featuring $HfO_2$ or $Al_2O_3$/$HfO_2$ bilayer, a LOR3A/S1805 bilayer was deposited and UV-exposed in agreement with the protocol previously described to define the top electrode geometry. On the other hand, due to the chemical incompatibility of $Al_2O_3$ and MF319 alkaline developer, on the $Al_2O_3$ samples a layer of AR-P 679.04 positive e-beam resist was spin-coated at 4000 RPM for 60 s (spin coating acceleration of 10000 RPM/s) and soft-baked at 160 °C for 60 s. The resist was then exposed via e-beam lithography (ZEISS, Ultra Plus Gemini provided with ELPHY MutiBeam Raith, acceleration voltage of 30 kV, gun aperture of 120 μm, beam current of 4.10 nA, dose of 330 μC cm$^{-2}$, step size of 600 nm) to define the top electrode geometry.

The electron-exposed samples were developed in AR 600-56 for 90 s with soft agitation to remove the exposed e-beam resist, then rinsed in IPA for 30 s to stop the development and dried with $N_2$. The resist-patterned samples were loaded into a thermal evaporator where a 10/55-nm-thick Ti/Au bilayer was deposited at a rate of 1.5 A/s at a residual chamber pressure of 2E-6 mbar. The deposited film was lifted-off in AR600-71 at 80 °C for 5 min with strong agitation, then rinsed in IPA for 30 s, and dried with $N_2$, to obtain the metallic top electrodes.

To check the impact of the lithography method on the electrical properties of the deposited insulators, top electrodes of MIM capacitors featuring an $HfO_2$ layer grown at 130 °C were also defined through electron beam-lithography using the following parameters:

1. Acceleration voltage of 10 kV, gun aperture of 120 μm, beam current of 3.78 nA, dose of 130 μC cm$^{-2}$, step size of 200 nm;

2. Acceleration voltage of 20 kV, gun aperture of 120 μm, beam current of 3.94 nA, dose of 250 μC cm$^{-2}$, step size of 400 nm;

3. Acceleration voltage of 30 kV, gun aperture of 120 μm, beam current of 4.10 nA, dose of 330 μC cm$^{-2}$, step size of 600 nm;

4. Acceleration voltage of 30 kV, gun aperture of 7.5 μm, beam current of 15.5 pA, dose of 330 μC cm$^{-2}$, step size of 5 nm;

All the fabricated samples were provided with bonding pads (200 × 500 μm × μm) and then used to connect the device with the chip carrier. Samples were glued using a small drop of AR-P 679.04, then left dry at RT for 1 hour on a 24-pin dual-in-line (DIL) chip carrier. Samples were bonded via wire wedge bonding (MP iBond5000 Wedge) using an Al/Si wire (1%, 25 μm wire diameter), leaving the user-bonder and the DIL chip carrier electrically connected to the ground.

**Electrical and Morphological Characterization of MIM Capacitors**

MIM capacitors were electrically measured in a [4]He closed-cycle cryostat (ICE OXFORD [DRY]ICE[3K]) with a measurement setup able to operate up to 1 MHz; the electrical characterization was performed at 300 K and 3K.

AC electrical characterization of MIM capacitors was performed measuring the capacitor impedance (magnitude and phase) in the frequency range from 100 Hz to 100 kHz (Precision LCR Meter, HP, 4284A), upon application of a sinusoidal voltage signal with 250 mV$_{rms}$ amplitude and 0 V bias.

DC electrical characterization of MIM capacitors was performed measuring the current-voltage curve upon application of an increasing voltage (SMU, Keithley, 2400, absolute voltage step = 100 mV, step delay = 1 s) and collecting the flowing current (TIA, FEMTO, DDPCA-300, gain=10$^{10}$ V/A, rise time = fast; Multimeter, Agilent, 34410A).

Morphological characterization of bottom and top electrode geometries was performed via optical microscopy (Leica, DM8000 M, provided with LEICA MC190 HD camera). On the other hand, Atomic Force Microscopy (AFM, Bruker, DIMENSION edge with ScanAsyst provided with an ASYLEC-01-R2 tip - silicon tip Ti/Ir coated, f$_0$=75 kHz, k=2.8 N/m - in tapping mode) was used to measure the insulators' thickness growth on a SiO$_2$(300-nm-thick)/Si sample. All the AFM photos were processed using Gwyddion software.

**Best-Fitting of Experimental Data**

Values of capacitance (C) and dielectric breakdown (V$_{BD}$) were extrapolated through best-fitting of experimental data. Best-fitting of impedance vs. frequency curves (magnitude and phase) was carried out using a capacitor lumped model described by (1) to retrieve C. The C vs. area curve was best-fitted with a linear curve. The area does not consider the area of the "vertical capacitor" formed with the bottom electrode thickness, since a total apport of < 0.2 % to the total capacitor area was estimated. Then, the C-axis intercept was subtracted to the retrieved C values to elide the systematic parasitic capacitors contribution. Best-fitting of current vs. voltage curves was carried out using a two-regions piecewise linear model to retrieve positive (V$_{BD}^+$) and negative (V$_{BD}^-$) breakdown voltages:

$$\begin{cases} I = I_1 + \sigma_1 V & if\ V < V_{DB} \\ I = I_1 + \sigma_1 V_{BD} + \sigma_2 (V - V_{BD}) & if\ V \geq V_{BD} \end{cases}$$

**Simulations of Electrons Trajectories and Average Deposited Energy in Electron Beam Lithography**

The simulations of electrons trajectories and average deposited energy in EBL were performed using ELPHY MultiBeam (Raith, NANOSUITE software release 6 SP 14.0). The substrate stack was reproduced to simulate the real case of study. From top to bottom, the following layers are involed:

- PMMA: t = 270 nm; atom. numb. = 3.2; atom. weig. = 6.674; dens. = 1.19 g cm$^{-3}$; ion. pot. = 0.074 keV.
- HfO$_2$: t = 30 nm; atom. numb. = 29.3; atom. weig. = 70.16; dens. = 9.68 g cm$^{-3}$; ion. pot. = 0.317 keV.
- Au: t = 40 nm; atom. numb. = 79; atom. weig. = 196.97; dens. = 19.32 g cm$^{-3}$; ion. pot. = 0.797 keV.
- Ti: t = 10 nm; atom. numb. = 22; atom. weig. = 47.867; dens. = 4.5 g cm$^{-3}$; ion. pot. = 0.0247 keV.
- SiO$_2$: t = 300 nm; atom. numb. = 10; atom. weig. = 20; dens. = 2.65 g cm$^{-3}$; ion. pot. = 0.135 keV.
- Si (substrate): t = bulk; atom. numb. = 14; atom. weig. = 28.0855; dens. = 2.33 g cm$^{-3}$; ion. pot. = 0.172 keV.

Then, Monte Carlo simulations were performed with $10^6$ electrons trajectories, point beam shape, and desired acceleration voltage.

**Statistical Analysis**

*Data presentation (e.g., mean ± RMSD)*

All the data referring to more than 3 devices are presented as mean value ($\bar{y}$) ± root mean square deviation (RMSD), where RMSD was evaluated as following:

$$RMSD = \sqrt{\frac{1}{n-1}\sum_{i=1}^{n}(y_i - \bar{y})^2}$$

*Sample size (n) for each statistical analysis*

All the data referring to more than 3 devices are presented indicating the sample size (n).

*Software used for analysis*

Best fitting and statistics of experimental data was performed using OriginLab OriginPRO 2024.